# Fiber-optic cascaded forward Brillouin scattering seeded by backward stimulated Brillouin scattering: conceptual proposal and experimental validation

Neisei Hayashi*[1], Yosuke Mizuno[2], Kentaro Nakamura[2], Sze Yun Set[1], and Shinji Yamashita[1]

[1] Research Center for Advanced Science and Technology, The University of Tokyo, 4-6-1, Komaba, Meguro-ku, Tokyo 153-8904, Japan

[2] Institute of Innovative Research, Tokyo Institute of Technology, 4259, Nagatsuta-cho, Midori-ku, Yokohama 226-8503, Japan

*Correspondence to: E-mail: hayashi@cntp.t.u-tokyo.ac.jp

We propose a method for generating cascaded forward Brillouin scattering (CFBS), based on a counter-propagated pump-probe technique, utilizing backward stimulated Brillouin scattering as its seed. The CFBS, induced by forward stimulated Brillouin scattering (FSBS), is generated via the energy transfer from the probe light to other acoustic resonance frequencies. Experimental results for the CFBS generated in a 390-m-long highly nonlinear fiber indicate that it has a high signal-to-noise ratio (SNR) and that the center frequencies of its acoustic resonance peaks agree with theoretical values. The sensing properties of the generated CFBS—specifically, the SNR dependence on pump and probe powers and the CFBS frequency shift dependence on temperature—were also verified.

Forward Brillouin scattering (FBS) is a nonlinear optical phenomenon that occurs in optical fibers and is generated by the interaction between incident light and acoustic waves propagating through the cross-sectional area of the fiber [1,2]. The propagating direction of FBS is the same as that of the incident light. FBS has been observed in various kinds of fibers, such as silica single-mode fibers (SMF) [1], polarization-maintaining fibers [3], and few-mode fibers [4]. When stimulated using co-propagated pump and probe lights [5,6], it is called forward stimulated Brillouin scattering (FSBS). In the FSBS process, if the pump power is relatively higher than that of the probe light and the center frequency of the probe light is the same as that of the first acoustic resonance peak of FBS, the energy of the pump can be transferred to other acoustic resonance frequencies. This phenomenon is called cascaded FBS (CFBS) [5,7,8]. The signal-to-noise ratio (SNR) of CFBS is reported to be higher than that of FBS [5].

To date, FBS has been used to develop megahertz-order digital logic devices [9], megahertz-order radio-frequency signal generators [10], temperature sensors [11], strain sensors [12], fiber-diameter sensors [13], and acoustic-impedance sensors [14-16]. Among them, acoustic-impedance sensing is a unique advantage of FBS-based sensors (the external acoustic impedance can be measured using only the linewidth of the FBS spectrum). If distributed acoustic-impedance sensing is achieved using conventional spatially resolving techniques [17-25], it could be exploited to implement distributed discriminative multi-parameter sensors for, for instance, cancer detection in the human body [26,27], underground oil layer detection [28], quality inspection of bioethanol [29], and NaCl concentration detection in water for cultivation of delicate tuna [30]. Furthermore, if transverse acoustic waves can be localized using a conventional correlation-domain technique in multi-core fibers, active distributed phase modulators [31] may be implemented. One of the major problems in achieving FBS-based distributed sensing is the propagating direction of the FBS, which prevents simple application of the conventional distributed Brillouin sensing techniques. To overcome this problem, in 2004, Tanaka et al. [32] proposed backward observation of FBS coupled with backward stimulated Brillouin scattering (BSBS), but the SNR of the backwardly observed FBS was extremely low because the Brillouin gain coefficient of FBS is much smaller than that of CFBS.

In this paper, first, we propose a method for generating CFBS seeded by BSBS using a counter-propagated pump-probe technique. In this method, the center frequency of the probe light is downshifted from that of the pump light by the sum of the Brillouin frequency shift (BFS) of BSBS and the first-order acoustic-resonance frequency shift of FBS. The CFBS induced by FSBS is generated via the energy transfer from the probe light to other acoustic resonance frequencies. Then, we present the results of experimental observation of the CFBS in a 390-m-long highly nonlinear fiber (HNLF) using BSBS as its seed. The measured CFBS has a high SNR and its center frequencies of the acoustic resonance peaks agree with theoretical values. Finally, we clarify the Brillouin properties of the generated CFBS peak at 941 MHz, such as the SNR dependence on pump and probe powers and the BFS dependence on temperature. As regards the SNR dependence on pump, the threshold of the CFBS is found to be 13.0 dBm with a −5.0-dBm probe light. As for the SNR dependence on probe, the SNR of the CFBS is optimized with a 17.0-dBm pump light and −5.0-dBm probe light. The temperature dependence coefficient of the BFS is measured to be 99 kHz/K, which is close to the theoretical value.

Various kinds of acoustic waves caused by thermal fluctuations exist in optical fibers [1].

One of these acoustic waves is the radial acoustic mode ($R_{0,m}$, where $m$ is the acoustic resonance mode), which literally induces displacements in the radial direction and also perturbs the refractive index in the fiber [1]. Optical scattering based on the $R_{0,m}$ mode is referred to as polarized FBS. The central frequency of the $m$-th acoustic mode ($\nu_{GB,m}$) is given by

$$\nu_{GB,m} = \frac{\upsilon y_m}{\pi d}, \qquad (1)$$

where $d$ is the fiber's outer diameter, $\upsilon$ is the velocity of the longitudinal acoustic waves, and $y_m$ is the value derived from the following equation [1]:

$$(1 - \alpha^2)J_0(y_m) - \alpha^2 J_2(y_m) = 0, \qquad (2)$$

where $\alpha$ is the shear acoustic velocity per longitudinal acoustic velocity, $J_0$ is the zero-order Bessel function, and $J_2$ is the second-order Bessel function. The central frequency is known to depend on temperature, with a dependence coefficient ($m = 11$) of 43.5 kHz/K in silica SMFs [33]. The coefficient of temperature dependence of the FBS spectrum is given by [11]:

$$\frac{d\nu_{GB,m}}{dT} = \left(\frac{1}{\upsilon}\frac{d\upsilon}{dT}\right)\nu_{GB,m} = K\nu_{GB,m}, \qquad (3)$$

where $K$ is a constant ($8.6\times10^{-5}$ /°C in silica SMFs) [11]. The $K$ value in the HNLF appears to be almost the same as that in a silica SMF, because the core/cladding materials of the HNLF and an SMF are similar to each other [6,11]. When the pump light and the probe light with center frequency shifted by the amount of acoustic resonance frequency from the pump light are injected into an optical fiber, the energy of the pump is transferred to the probe light, which enhances the probe light. This phenomenon is called FSBS [6]. The pump power of FSBS is transferred to a higher Stokes order [5,8] when the pump power is higher than that of the probe power and the center frequency of the probe light is congruent with that of the first acoustic resonance frequency. This cascading process is unique to FBS, although the cascaded optical fields coherently drive the same phonon field producing successive parametric frequency shifts [5].

The basic concept underlying the principle of the CFBS generation seeded by BSBS is depicted in Fig. 1. When we inject pump light with a center frequency of $\nu_0$ and probe light with a center frequency of $\nu_0 - \nu_B + \nu_{GB,1}$ (see Figs. 1(a)(b)), the energy of the pump light is transferred to the probe light (see Fig. 1(c)) [34] because the BSBS linewidth is sufficiently broad. Then, FBS is generated by the enhanced probe (see Fig. 1(d)) [9], and the energy of the pump is further transferred to the acoustic resonance peak of the FBS and probe light (see Fig.

1(e)). Subsequently, this FBS signal and spontaneous backward Brillouin scattering generated by the pump light are enhanced by the pump light (see Fig. 1(f)) [5,7], which generates BSBS for seeding CFBS (see Fig. 1(g)). Finally, the energy of the probe light is transferred to the seed and CFBS is generated [8] (see Fig. 1(h)).

In experiment, we employed an ultraviolet curable coated 390-m-long silica HNLF (HNDS1614CA-4-3-3, Sumitomo) as the fiber under test (FUT). This HNLF had a core diameter of 3.5 μm, a cladding diameter of 117.0 μm, a core refractive index of ~1.46, a dispersion-shifted structure, and a propagation loss of ~0.76 dB/km at 1.55 μm. The BFS was 9.092 GHz at 1.55 μm, and the Brillouin threshold power was 13 dBm with the pump-probe technique. The schematic setup used to observe the CFBS spectrum is shown in Fig. 2. It is fundamentally the same as that previously reported [9]. All the optical paths excluding the FUT are standard silica SMFs. The output from a distributed-feedback laser diode at 1546 nm (linewidth: 1.2 kHz, output power: <6 dBm) was divided into two: probe and pump lights. The center frequency of the probe light was shifted by several gigahertz with a single-sideband modulator (SSBM; carrier suppression ratio: >20 dB). This value well covers the BFS ($v_B$) of backward SBS and the first acoustic resonance frequency of FBS. After power amplification using an erbium-doped fiber amplifier (EDFA), the probe light was injected into the FUT. The pump light was injected into the FUT from the other end, after amplified using another EDFA. The CFBS light was guided to a photodetector (PD) along with the transmitted probe light via a variable optical attenuator (VOA). Using the PD, the beat signal between the CFBS and the transmitted probe light was converted into an electrical signal, which was observed using an electrical spectrum analyzer (ESA). The polarization state in the FUT was averaged with a polarization scrambler (PSCR). The amplified spontaneous emission noise from each EDFA was suppressed using two optical filters (OF) with a narrow bandwidth (~10 GHz). The ambient temperature was 27ºC.

Figures 3(a)(b) show the CFBS spectrum observed for a pump power of 13.9 dBm and a probe power of −5.0 dBm. The attenuation of the VOA was set to 8.7 dB. The horizontal axis was normalized by subtracting the electrical noise of the PD and the ESA as the noise floor from the measured spectrum. When the center frequency of the probe light was set to 9.092 GHz, a small peak was observed at 400 MHz, which originates from the second Brillouin peak (see Fig. 3 (a)). Further, when the center frequency of the probe was set to 9.072 GHz, approximately 20 clear peaks appeared in the frequency range from 20 MHz to 1000 MHz, which agrees with the theoretical center frequencies of radial acoustic resonance peaks. This theoretical $v_{\text{GB},m}$ value was calculated by substituting $v = 5590$ m/s (6) and $\alpha = 0.624$ [11] into Eq. (1) (see Fig. 3 (b)). The frequency range in which acoustic resonance peaks were observed was approximately three times broader than that of the small-core photonic crystal fibers [5]. The CFBS-BFS values of each peak agreed well with the theoretical values of each $R_{0,m}$ mode [1].

Next, we measured the pump power dependence of the CFBS spectrum near the peak at ~941 MHz (corresponding to the $R_{0,20}$ mode, which exhibited one of the clearest peaks in the higher frequency section). The probe power was −5.0 dBm. The spectral peak power increased with the pump power, but the noise floor also increased (Fig. 4(a)). Therefore, here we define the SNR as the ratio between the spectral peak power and the spectral power at 961 MHz. Figure 4(b) shows the pump power dependence of the CFBS spectrum, which was normalized so that the spectral power at 961 MHz was 0 dB. The spectrum at the pump power of 10.0 dBm (regarded as the noise floor of the ESA) was subtracted from all the spectra. The SNR increased with increasing pump power as shown in Fig. 4(c). The threshold of the CFBS was ~13.0 dBm, which was the almost the same as that of BSBS. Then, the SNR was limited to 4.4 dB from 17.0-dBm pump power by limiting the maximum power of the EDFA in this experiment.

Subsequently, we measured the probe power dependence of the CFBS spectrum at ~941 MHz. The pump power was 17.0 dBm. As shown in Fig. 5(a), as the probe power increased, the spectral peak power and the noise floor increased. The probe power dependence of the CFBS spectrum normalized using the spectral power at 961 MHz is shown in Fig. 5(b). The spectrum at the probe power of −22.0 dBm was subtracted from all the spectra in order to suppress the influence of the noise of the ESA. The SNR became maximal at a probe power of −5.0 dBm. The SNR dependence on the probe power is shown in Fig. 5(c). The SNR was almost linearly increased up to −5.0 dBm. The highest SNR was measured to be 4.4 dB when the pump power was 17 dBm and the probe power was −5.0 dBm. The probe power was limited by the input power of the PD in this experiment.

Finally, we measured the temperature dependence of the CFBS spectrum. The temperature of the FUT was changed using a thermostatic chamber from 30°C to 70°C. The spectrum near the peak at ~941 MHz dependence on temperature is shown in Fig. 6(a). The center frequency of the spectrum was shifted to a higher frequency with increasing temperature. The power of the spectrum decreased for the temperature range 30°C to 70°C. The temperature dependence of the center frequency shift of the spectra is shown in Fig. 6(b). The dependence was linear with a coefficient of 99 kHz/K, which is close to the value calculated using Eq. (3).

The slight difference between the measured value and the theoretical value originates from the material properties [35]. This coefficient is clearly different from that of BSBS in the HNLF [36]. The spectra normalized in order for the power at 961 MHz to be 0 dB is shown in Fig. 6(c). The SNR decreased with increasing temperature, as shown in Fig. 6(d). The SNR remained almost the same between 30°C and 40°C, but then decreased linearly from 40°C to 70°C. Although the BFS of CFBS and that of BSBS were both linearly dependent on temperature, this nonlinear SNR dependence on temperature was observed, probably because of the Lorentzian shape of the BSBS spectrum used as seed. If the center frequency of the probe light is fixed at only the first acoustic resonance frequency, the temperature measurement range is limited. For wide-range measurement of temperature, optimization of the center frequency of the probe light is required. This information will be useful for discriminating temperature and other parameters in BSBS-seeded CFBS-based sensing systems in the future.

In summary, this paper proposed a method for generating BSBS-seeded CFBS using a counter-propagated pump-probe technique. Experimental results using the 390-m-long HNLF indicate that the CFBS had a relatively high SNR and that the center frequencies of its acoustic resonance peaks agreed with theoretical values. The Brillouin properties of the generated CFBS, such as the SNR dependence on pump power, the SNR dependence on probe power, and the BFS dependence on temperature, were then clarified. As for the SNR dependence on pump power, the threshold of CFBS was measured to be 13.0 dBm with a −5.0-dBm probe light, which was almost identical to that of BSBS. In contrast, as for the SNR dependence on probe power, the SNR of the CFBS became maximal with a 17.0-dBm pump light and a −5.0-dBm probe light. The temperature dependence coefficient of the CFBS-BFS values was measured to be 99 kHz/K, which was close to the theoretical value. We believe that these findings will aid the development of CFBS-based distributed acoustic-impedance sensors for underground oil layer detection, cancer detection in the human body, etc., in the near future.

**Acknowledgments**

This study was financially supported by JSPS KAKENHI (17H04930, JP16H00902, 17J07226, 25009080, 17K14692) and by research grants from the Japan Gas Association, the ESPEC Foundation for Global Environment Research and Technology, the Association for Disaster Prevention Research, the Fujikura Foundation, and the Japan Association for Chemical Innovation.

**References**

1. R. M. Shelby, M. D. Levenson, and P. W. Bayer, Phys. Rev. B **31**, 5244 (1985).
2. A. J. Poustie, J. Opt. Soc. Am. B **10**, 691 (1993).
3. N. Nishizawa, S. Kume, M. Mori, T. Goto, and A. Miyauchi, Opt. Rev. **3**, 29 (2001).
4. T. Matsui, K. Nakajima, and F. Yamamoto, Appl. Opt. **54**, 6093 (2015).
5. M. S. Kang, A. Nazarkin, A. Brenn, and P. S. J. Russell, Nat. Phys. **5**, 276 (2009).
6. J. Wang, Y. Zhu, R. Zhang, and D. J. Gauthier, Opt. Exp. **19**, 5339 (2011).
7. E. A. Kittlaus, H. Shin, and P. T. Rakich, Nat. Photon. **10**, 463 (2016).
8. C. Wolff, B. Stiller, B. J. Eggleton, M. J. Steel, and C. G. Poulton, New J. Phys. **19**, 023021 (2017).
9. N. Hayashi, H. Lee, Y. Mizuno, and K. Nakamura, IEEE Photon. J. **8**, 7100707 (2016).
10. Y. London, H. H. Diamandi, and A. Zadok, APL Photon. **2**, 041303 (2017).
11. Y. Tanaka and K. Ogusu, IEEE Photon. Technol. Lett. **10**, 1769 (1998).
12. Y. Tanaka and K. Ogusu, IEEE Photon. Technol. Lett. **11**, 865 (1999).
13. M. Ohashi, S. Naotaka, and S. Kazuyki, Electron. Lett. **28**, 900 (1992).
14. Y. Antman, A. Clain, Y. London, and A. Zadok, Optica **3**, 510 (2016).
15. N. Hayashi, Y. Mizuno, K. Nakamura, S. Y. Set, and S. Yamashita, Opt. Exp. **25**, 2239 (2017).
16. D. M. Chow, M. A. Soto, and L. Thévenaz, Proc. SPIE **10323**, 1032311 (2017).
17. T. Horiguchi and M. Tateda, J. Lightw. Technol. **7**, 1170 (1989).
18. J. Urricelqui, M. Sagues, and A. Loayssa, Opt. Exp. **22**, 17403 (1989).
19. Y. Dong, L. Teng, P. Tong, T. Jiang, H. Zhang, T. Zhu, L. Chen, X. Bao, and Z. Lu, Opt. Lett. **40**, 5003 (2015).
20. Y. Dong, D. Ba, T. Jiang, D. Zhou, H. Zhang, C. Zhu, Z. Lu, H. Li, L. Chen, and X. Bao, IEEE Photon. J. **5**, 2600407 (2013).
21. T. Kurashima, T. Horiguchi, H. Izumita, S. Furukawa, and Y. Koyamada, IEICE Trans. Commun. **E76-B**, 382 (2014).
22. L. E. Y. Herrera, G. C. Amaral, and J. P. von der Weid, Appl. Opt. **55**, 1177 (2016).
23. K. Hotate and T. Hasegawa, IEICE Trans. Electron. **E83-C**, 405 (2000).
24. R. K. Yamashita, W. Zou, Z. He, and K. Hotate, IEEE Photon. Technol. Lett. **24**, 1006 (2012).
25. Y. Mizuno, W. Zou, Z. He, and K. Hotate, Opt. Exp. **16**, 12148 (2008).
26. M. L. Palmeri, M. H. Wang, N. C. Rouze, M. F. Abdelmalek, C. D. Guy, B. Moser, A.

M. Diehl, and K. R. Nightingale, J. Hepatol. **55**, 666 (2011).

[27.] Y. Kato, Y. Wada, Y. Mizuno, and K. Nakamura, Jpn. J. Appl. Phys. **53**, 07KF05 (2014).

[28.] Schlumberger, Ltd., Log Interpretation Principles/Applications (Schlumberger Educational Services, Texas, 1991).

[29.] S. K. Srivastava, R. Verma, and B. D. Gupta, Sens. Actuat. B Chem. **153**, 194 (2011).

[30.] D. Ma, Q. Ding, D. Li, and L. Zhao, Sens. Lett. **8**, 109 (2010).

[31.] H. H. Diamandi, Y. London, and A. Zadok, Optica, **4**, 289 (2017).

[32.] Y. Tanaka, H. Yoshida, and T. Kurokawa, Meas. Sci. Technol. **15**, 1458 (2004).

[33.] Y. Antman, Y. London, and A. Zadok, Proc. SPIE **9634**, 96345C (2015).

[34.] G.P. Agrawal, *Nonlinear fiber optics* (Academic Press, California, 1995).

[35.] N. Hayashi, K. Suzuki, S. Y. Set, and S. Yamashita, Appl. Phys. Exp. **10**, 092501 (2017).

[36.] M.R. Lorenzen, D. Noordegraaf, C.V. Nielsen, O. Odgaard, L. Gruner-Nielsen, and K. Rottwitt, Electron. Lett, **45**, 125 (2009).

**Figure Captions**

Fig. 1. Generation process of CFBS using BSBS as seed. (a) Schematic setup. (b) Initial state of pump and probe light. (c) Energy transfer from pump to probe light. (d) FBS generation by enhanced probe. (e) Enhancement of first resonance FBS peak and probe. (f) Generation of BSBS as seed of CFBS. (g) Energy transfer from probe to the seed. (h) Generation of CFBS using BSBS as seed.

Fig. 2. Schematic setup used to observe the cascaded forward Brillouin scattering (CFBS) using backward stimulated Brillouin scattering (BSBS) as seed. EDFA: erbium-doped fiber amplifier, ESA: electrical spectrum analyzer, FG: function generator, HNLF: highly nonlinear fiber, OF: optical filter, PD: photodetector, PSCR: polarization scrambler, SSBM: single-sideband frequency modulator; VOA: variable optical attenuator.

Fig. 3. (a) Wide-range view of the spectrum when the center frequency of the probe light was set to the Brillouin frequency shift (BFS). (b) Wide-range view of the observed CFBS spectrum (red), and theoretical center frequencies of $R_{0,m}$ (blue).

Fig. 4. (a) Pump power dependence of the CFBS spectrum around 941 MHz. (b) Pump power dependence of the CFBS spectrum normalized in order for the spectral power at 961 MHz to be 0 dB. (c) Pump power dependence of the signal-to-noise ratio (SNR) of the CFBS measurement.

Fig. 5. (a) Probe power dependence of the CFBS spectrum around 941 MHz. (b) Probe power dependence of the CFBS spectrum normalized in order for the spectral power at 961 MHz to be 0 dB. (c) Probe power dependence of the SNR of the CFBS measurement for a pump power of 17.0 dBm.

Fig. 6. (a) Temperature dependence of the CFBS spectrum around 941 MHz. (b) Temperature dependence of the center frequency shift of the spectrum. The solid line is a linear fit. (c) Temperature dependence of the CFBS spectrum normalized in order for the spectral power at 961 MHz to be 0 dB. (d) Temperature dependence of the SNR.

**Figures**

Fig. 1

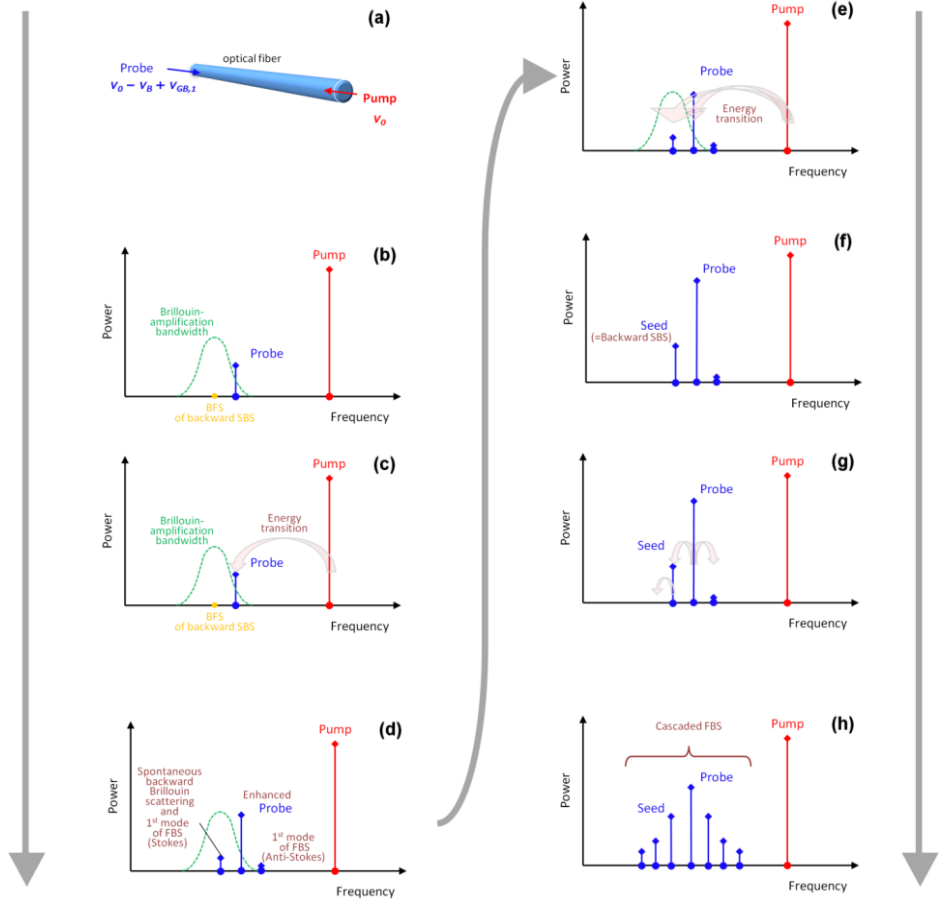

Fig. 2

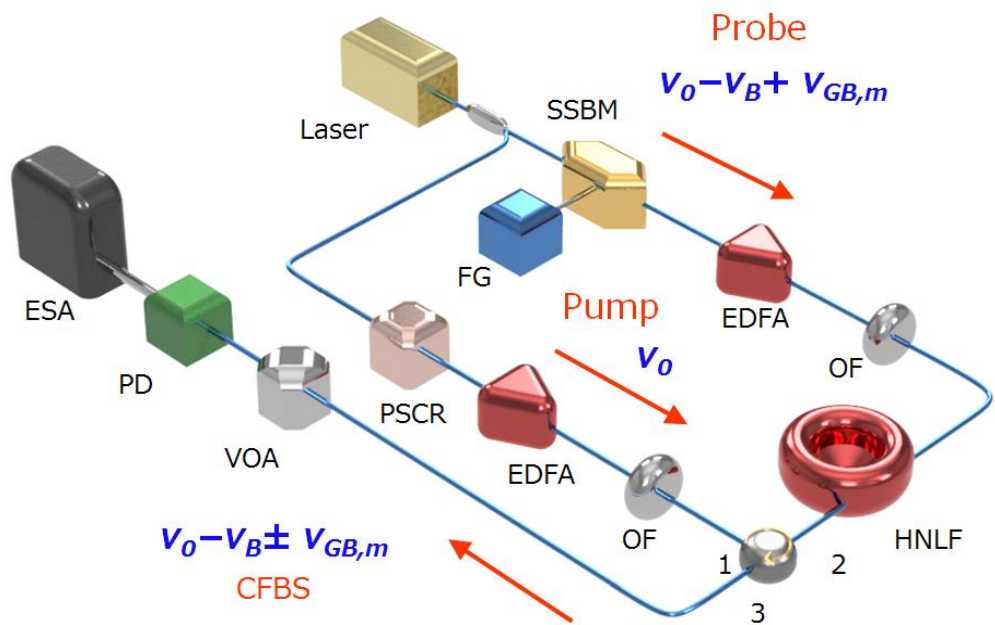

Fig. 3

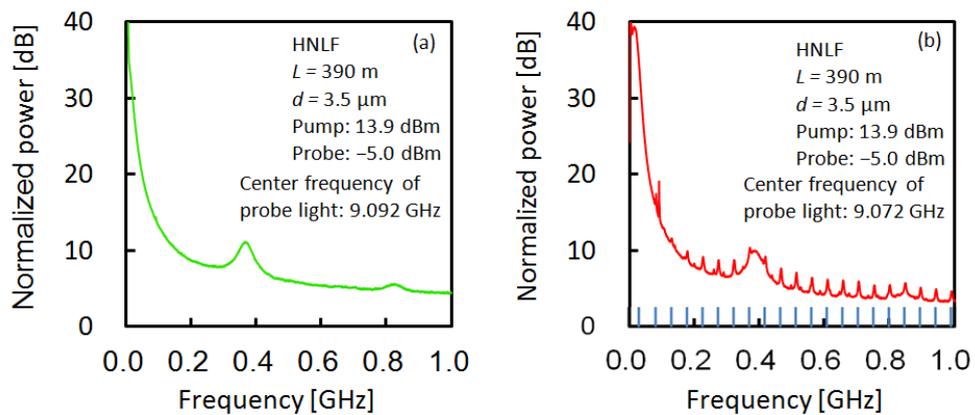

Fig. 4

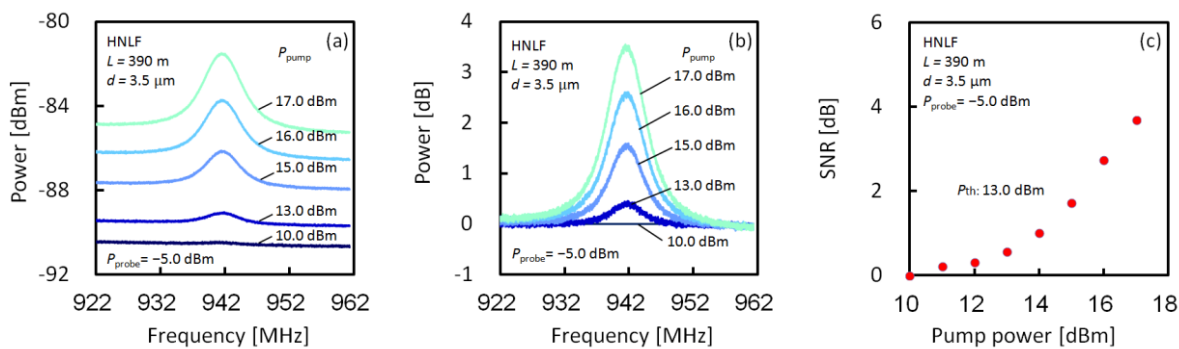

Fig. 5

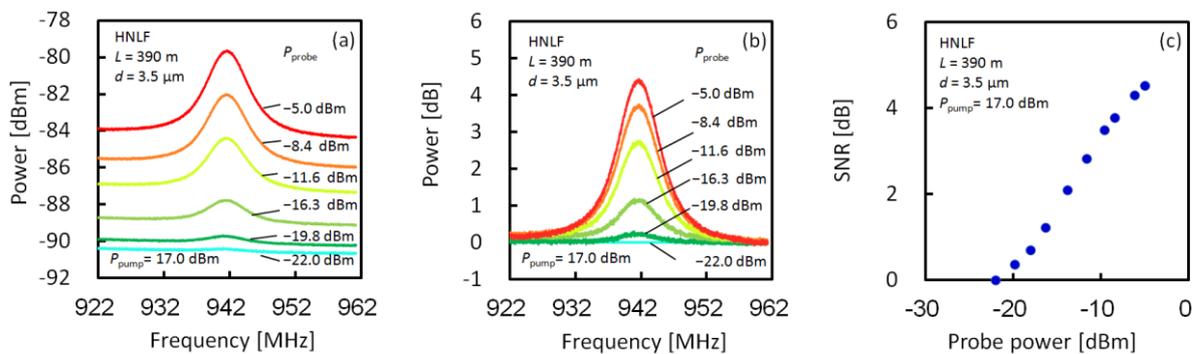

Fig. 6

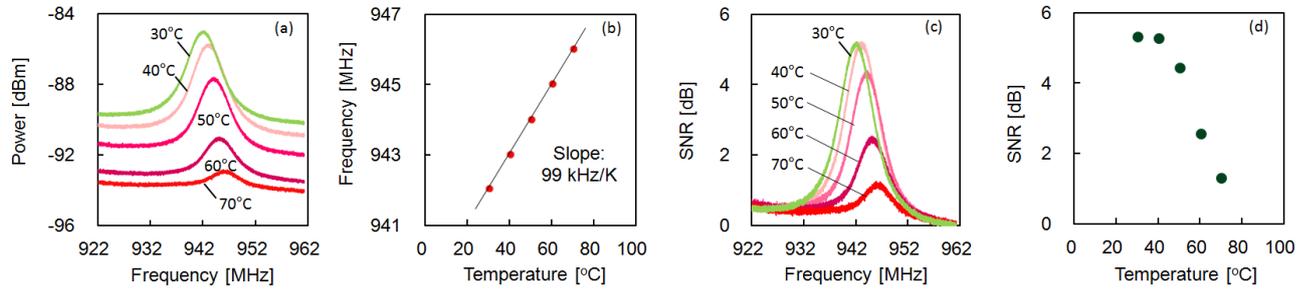